\documentclass[preprint]{sig-alternate-05-2015}

\makeatletter
\def\@copyrightspace{\relax}
\makeatother
\begin{document}






%

\title{LiRa: A New Likelihood-Based Similarity Score For Collaborative Filtering
}
%
%
%
%
%

\numberofauthors{5} 
%
\author{
%
%
\alignauthor
Veronika Strnadov\'a-Neeley\\
       \affaddr{University of California, Santa Barbara}\\
\alignauthor
Ayd\i n Bulu\c{c}\\
       \affaddr{Lawrence Berkeley National Lab}\\
\alignauthor 
John R. Gilbert\\
       \affaddr{University of California, Santa Barbara}\\
\and  
\alignauthor Leonid Oliker\\
       \affaddr{Lawrence Berkeley National Lab}\\
\alignauthor Weimin Ouyang\\
       \affaddr{University of California, Santa Barbara}\\
}

\maketitle
\begin{abstract}
Recommender system data presents unique challenges to the data mining, 
machine learning, and algorithms communities. The high missing data rate, 
in combination with the large scale and high dimensionality
typical of recommender systems data, requires new tools and methods for 
efficient data analysis. Here, we address the challenge of evaluating 
similarity between users in a recommender system, where for each 
user only a small set of ratings is available. We present a new similarity score, 
that we call LiRa, based on a statistical model of 
user similarity for large-scale, discrete valued data with many 
missing values. We show that this likelihood ratio-based score
is more effective at identifying similar users than traditional 
similarity scores in user-based collaborative filtering, 
such as the Pearson correlation coefficient. We argue that our approach 
has significant potential to improve both accuracy and scalability 
in collaborative filtering.
\end{abstract}

%
%


%
%

%
%



\section{Introduction}
\label{sec:intro}
Recommender systems arose as a way to provide personalized recommendations,
in a setting where the number of available items, products, or options
to a user is too large to sift through manually.  
Collaborative filtering has proven to be an effective 
approach for recommendation, relying on the 
similarity of users or items in a system to predict 
future user preferences.
The premise underlying user-based collaborative filtering is that
similar users tend to rate items similarly, therefore 
to predict how a user $u$ will rate an item $i$, 
we should look at the ratings given to $i$ by users similar to $u$.
In item-based collaborative filtering,
the assumption is that similar items
tend to be rated similarly by the users, 
therefore the rating prediction should be based
on the ratings given by user $u$ 
to items similar to $i$.
A central component of any collaborative 
filtering algorithm is the choice of similarity score 
that is used to evaluate user-user or item-item similarity. 
\begin{figure}
  \includegraphics[scale=0.45]{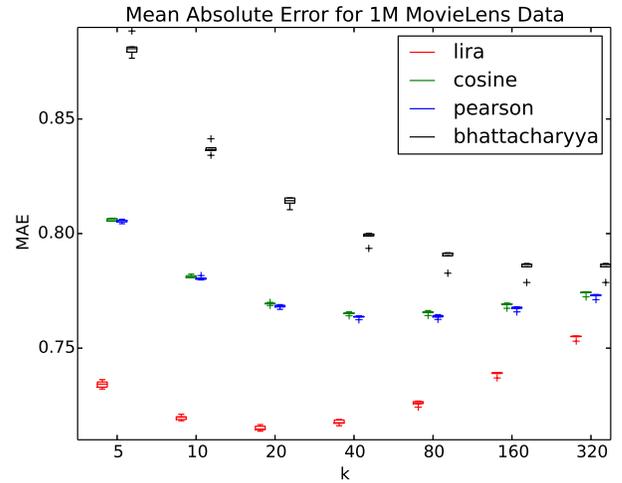}
  \caption{Prediction accuracy of the kNN method using several similarity scores on the 1M MovieLens dataset. The LiRa similarity attains the lowest MAE across tested values of $k$. The  
  greater difference between LiRa and other scores at lower values of $k$ indicate its better ability to find 
  similar users.}
  \label{fig:mae_1Mdataset}
\end{figure}

Despite many improvements and successes of modern model-based 
collaborative filtering, the k-Nearest-Neighbor (kNN) method remains
a popular and widely used approach, in large part due to its
simplicity and scalability.\cite{desrosiers2011comprehensive}
To perform user-based collaborative filtering, the kNN method predicts
the rating $p_{ui}$ for an item $i$ by a user $u$ by selecting
the $k$ most similar users to user $u$ who have rated item $i$.
The prediction $p_{ui}$ is then
computed by averaging the ratings given to item $i$ by these $k$ similar users.
Similarly, in item-based kNN, the ratings of the $k$ items 
most similar to item $i$ and rated by user $u$ are used to 
compute $p_{ui}$.
Many variants of kNN have been proposed and investigated in order
to provide guidelines for optimal parameter settings and 
implementation choices. The choice of similarity score
has consistently been shown to be highly influential on 
rating prediction accuracy. \cite{desrosiers2011comprehensive, herlocker2002knnempirical, patra2015new, ahn2008new}

In this work, we present a new similarity score, called LiRa, 
for large-scale, 
discrete-valued and high-dimensional data with many missing values.
We use an empirical evaluation on real data to show its effectiveness
in finding similar users in user-based collaborative filtering.
We also present an evaluation of several similarity scores' ability 
to detect clustered points in synthetic data sets,
revealing fundamental properties of these scores 
that are important in their application to recommender system data.

\section{Motivation and Background}
\label{sec:background}
Our work stems from a well-known problem in collaborative filtering:
RS data is often very sparse, meaning that 
in a system with $m$ users and $n$ items, the number of
ratings observed is typically much less than the $mn$ user-item pairs.  
Thus in approaches that seek to predict future ratings based
on user-user or item-item similarity, it is important to 
consider how the sparsity of ratings affects the similarity score.

Popular choices of similarity scores for user-based collaborative
filtering 
include the Pearson correlation coefficient and the cosine similarity,
which are ``commonly accepted as the best choice."\cite{ricci2011introduction} 
We review these traditional scores briefly to highlight
the core issue that will be addressed with our new similarity score,
presented in Section \ref{sec:liradef}.

For two users $u$ and $v$, 
let $I_{uv}$ be the set of 
co-rated items, i.e. those items that were rated by both $u$
and $v$. Let $r_{ui}$ be the rating given to item $i$ by user $u$. 
Then, the Pearson correlation between users $u$ and $v$, $PC(u,v)$,
is defined as follows \cite{ricci2011introduction}:
\begin{equation} 
PC(u,v) = \frac{\sum_{i \in I_{uv}} (r_{ui} - \bar{r}_u)(r_{vi} - \bar{r}_v) }{\sqrt{ \sum_{i \in I_{uv}} (r_{ui} - \bar{r}_u)^2 \sum_{i \in I_{uv}} (r_{vi} - \bar{r}_v)^2 } } 
\end{equation}
where $\bar{r}_u$ is the average rating given by user $u$
to the items in $I_{uv}$, and similarly for $\bar{r}_v$.
The Pearson correlation is a measure of linear correlation between
user $u$ and $v$'s ratings, and takes on values between -1 and 1.

The Cosine Vector similarity between users $u$ and $v$, 
or $CV(u,v)$, is a measure of the angle between the $N$-dimensional 
vectors defined by user $u$ and $v$'s ratings. More specifically,
if we let $y_u \in R^N$ be the vector with $y_{ui} = r_{ui}$ 
for rated item $i$, and $y_{ui} =0$ otherwise,
then:
\begin{equation}
 CV(u,v) = \frac{y_u^T y_v}{||y_u|| ||y_v ||} = \frac{\sum_{i \in I_{uv}} r_{ui}r_{vi} }{ \sqrt{ \sum_{i \in I_{u} } r_{ui}^2  \sum_{j \in I_{v} } r_{vj}^2 } }
\end{equation}
where $I_u$, $I_v$ are the sets of items rated by $u$ and $v$, respectively.
The cosine of the angle between two vectors ranges from -1 to 1, with
 1 indicating perfectly matching entries in
 both vectors.
 A CV similarity of 1, therefore, indicates perfectly
 matching entries.
However, although a cosine of 0 indicates
orthogonal vectors in an $N$-dimensional vector space,
the cosine of two rating vectors will only be 0 if
the there are no co-rated items in raw (unnormalized)
data. 

A major issue with both of these scores is their
lack of consideration for missing data. 
Although the PC similarity, which is equivalently 
the sample Pearson correlation coefficient,
is a consistent estimator of the population
correlation coefficient for large sample sizes, 
the number of co-rated items between $u$ and $v$, 
or $|I_{uv}|$, is often so small that the PC similarity is not
reliable.

Similarly, we can think of the Cosine Vector similarity as 
the cosine of the angle between two vectors that represent the 
projection of user ratings onto the space spanned by
the $|I_{u} \cup I_{v}|$ dimensions in which data is observed,
but the value of this angle in higher dimensions has is treated the 
same as its value in lower dimensions.

We conclude this explanation of the drawbacks of popular
similarity scores used on RS data with the following example.
Suppose we want to compute the similarity between two users
represented by the following identical rating vectors, 
constructed by following the definition of vectors used by
the CV similarity: 
\begingroup\makeatletter\def\f@size{8}\check@mathfonts
\begin{equation}
\label{eq:vecs1}
y_u = \left[ \begin{array}{cccccc} 1 & 1 &  0 & 0 & 0 & 2 \end{array} \right], \\
y_v = \left[ \begin{array}{cccccc} 1 & 1 &  0 & 0 & 0 & 2 \end{array} \right] 
\end{equation}
\endgroup
Both CV and Pearson will yield a score of 1 between $u$ and $v$. 
If we increase the amount of data observed, leaving the vectors
identical:
\begingroup\makeatletter\def\f@size{8}\check@mathfonts
\begin{equation}
x_u = \left[ \begin{array}{cccccc} 1 & 1 &  5 & 4 & 4 & 2 \end{array} \right] 
x_v = \left[ \begin{array}{cccccc} 1 & 1 &  5 & 4 & 4 & 2 \end{array} \right] 
\end{equation}
\endgroup
the CV and Pearson similarities will remain the same.

However, we have observed twice as much data in the second
case -- shouldn't our similarity score reflect more 
confidence in the similarity computation as we increase
the amount of data we have available for input? This question
motivated us to derive a new similarity score, which 
we refer to as the Likelihood Ratio, or LiRa, similarity.

\section{The Likelihood Ratio Similarity}
\label{sec:liradef}
The idea to use a likelihood-based score for similarity computations
in RS data was inspired in part by the \textit{LOD score}
popular in genetic mapping~\cite{cheema2009computational}
and the concept of \textit{modularity} in community
detection~\cite{newman2006modularity}.
In both cases, the concept of similarity is based on comparing
the likelihood of the observed data, under some assumptions
on the underlying data structure, to the likelihood of
observing the data by chance. In genetic mapping,
the LOD score relates the likelihood of observing
genetic marker data, assuming genetic linkage,
to the likelihood of observing the same genotypes by chance. 
Newman~\cite{newman2006modularity} introduced the idea
that a community structure contains many more edges 
than expected if the edges among social network vertices 
were generated at random.
Extending these ideas to the RS domain, we present the Likelihood Ratio
Similarity.
\subsection{Definition of the LiRa Similarity}
For two discrete-valued vectors $x_u$ and $x_v$,
we define the Likelihood Ratio (LiRa) Similarity as follows:
\begingroup\makeatletter\def\f@size{8}\check@mathfonts
\begin{equation}
\textnormal{LiRa}(x_u,x_v) = \log_{10} \frac{ p(\textnormal{differences in } x_u \textnormal{ and } x_v \textnormal{| same cluster  
}) }{ p(\textnormal{differences in }x_u \textnormal{ and } x_v | \textnormal{ pure chance}) }
\end{equation}
\endgroup
where the numerator in the ratio
is the probability of
observing the values in $x_u$ and $x_v$, 
assuming $x_u$ and $x_v$ belong to the 
same cluster in our cluster model, and
the probability in the denominator is
set by assuming that 
the entries in $x_u$ and $x_v$ were generated 
uniformly at random.

Suppose that the entries in each vector can take on only a finite
number $d$ of discrete values $\mathcal{V} = \lbrace 1, 2, \dots, d \rbrace$.
Then, we can easily compute the probability that we observe
the values $x_{ui}$ and $x_{vi}$ for co-observed entry $i$ by chance,
assuming that the values are generated uniformly and independently
at random. 
This is probability is simply $\frac{1}{d^2}$. Therefore, the probability
that the two vectors match exactly in a particular entry $i$
is $p(|x_{ui} - x_{vi}|=0) = \frac{d}{d^2} = \frac{1}{d}$.
Similarly, we can easily derive $p(|x_{ui} - x_{vi}|=\delta)$
for $\delta = 1, ..., d-1$. The denominator in the ratio in LiRa
is thus defined as: 
\begin{equation} p(\textnormal{differences in }x_u \textnormal{ and } x_v | \textnormal{ pure chance}) = \prod_{\delta=0}^{d-1} b_\delta^{\#\delta}  
\end{equation}
where $ b_{\delta} = p(|x_{ui} - x_{vi}|=\delta) $,
assuming that $x_{ui}$ and $x_{vi}$ were generated by 
a uniform distribution over the values $\mathcal{V}$,
and $\#\delta$ is the number of times that we observe
a difference of $\delta$ in the co-observed entries.

The challenge is to 
define the probability of observing a difference of $\delta$ in
the values $x_{ui}$ and $x_{vi}$,
under the assumption that $x_u$ and $x_v$ belong to the same cluster.  
This is not a trivial task, and we argue that
this model will be dependent on the application of interest.
For RS data, 
we make two assumptions
that we believe lead to one plausible model: 
(1) An underlying cluster structure exists in RS data: 
There exist a set of clusters $\mathcal{C}_1, ..., \mathcal{C}_{\kappa}$
such that each user $u$ belongs to at least one $\mathcal{C}_{c}$, and 
(2) The probability distribution on differences in user ratings 
is fixed within a cluster, with a greater probability of observing
matching than mismatched ratings.
These assumptions encapsulate the intuition that 
similar users tend to rate items similarly.

With these assumptions, we define the following probability
distribution over the differences $|x_{ui} - x_{vi}|$:
\begin{equation}
c_\delta = p(|x_{ui} - x_{vi}| = \delta | \textnormal{ same cluster}) = \left( \frac{1}{2} \right)^{\delta+1}
\end{equation} 
with the exception that 
\begin{equation}
c_{d-1} = p(|x_{ui} - x_{vi}| = d-1) = 1 - \sum_{\delta = 0}^{d-2} c_\delta = \frac{1}{2^{d -1}}
\end{equation}
to ensure a proper probability distribution.
Therefore the numerator in the ratio in LiRa becomes:
\begin{equation}
  p(\textnormal{differences in }x_u \textnormal{ and } x_v | \textnormal{ same cluster }) = \prod_{\delta=0}^{d-1} c_\delta^{\#\delta}   
\end{equation}
where $c_{\delta}$ and $\#\delta$ are defined above, and
$x_{ui} = r_{ui}$ if user $u$ rated item $i$, and
$-$ otherwise,
where $-$ indicates a missing value.

We emphasize that both $x_u$ and $x_v$ may have many missing
values, which are not taken into account when evaluating
these probabilities. In particular, the values are not simply treated
as 0's as in the Cosine Vector similarity score.
On the other hand, as long as $\frac{1}{2}>\frac{1}{d}$,
 the LiRa score increases with a greater number of 
 matching co-observed entries, and 
 in general the contribution to the LiRa score
 of the rating difference
 for a co-observed item $i$ will depend on the
 number of discrete values $d$.
 For example, with $d=5$, $b_1>c_1$, but at $d=10$,
 $b_1<c_1$, thus a difference of 1 in a rating scale
 of 1 to 5 will decrease the LiRa score,
 whereas on a rating scale of 1 to 10, a difference of
 1 in user ratings will increase it.
 To see this, notice we can re-write the LiRa score as:
\begin{equation} 
\sum_{d=1}^\delta (\#\delta)\log_{10} \left( \frac{c_\delta}{b_\delta}\right) 
\end{equation}
 and thus $\log_{10}\left(c_\delta/b_\delta\right)$ is the 
 amount that a pair of co-observed
 ratings $x_{ui}$ and $x_{vi}$ with a rating difference
 of $\delta$ will contribute to the
 similarity score. 

In future work, we plan to explore other, perhaps more plausible, multinomial probability distributions
over the differences in user ratings that capture the intuition
that users in the same cluster should rate items with very close rating values.
However, we claim that this simple model 
captures enough of the intuition that 
users with similar preferences are more likely
to agree than disagree in their ratings of
the same item. We will show in Section \ref{sec:empiricalevaluation} that 
these assumptions lead
to a useful similarity score for RS data. 
 
We conclude with an example using the same the vectors $y_u$ and $y_v$ 
from Equation \ref{eq:vecs1}.
The corresponding vectors $x_u$ and $x_v$ are:
\begingroup\makeatletter\def\f@size{8}\check@mathfonts
\[ x_u = \left[ \begin{array}{cccccc} 1 & 1 &  - & - & - & 2 \end{array} \right], \\
x_v = \left[ \begin{array}{cccccc} 1 & 1 &  - & - & - & 2 \end{array} \right] \]
\endgroup
Suppose that there are $d=5$ discrete
rating values in the data set. We get the LiRa similarity:
\begin{equation}
\textnormal{LiRa}(x_u, x_v) = \log_{10} \frac{ (\frac{1}{2})^3 }{ (\frac{1}{5})^3 } = 1.19
\end{equation}
Now consider LiRa$(x_u, x_v)$ when we observe the full vectors:
\begingroup\makeatletter\def\f@size{8}\check@mathfonts
\[x_u = \left[ \begin{array}{cccccc} 1 & 1 &  5 & 4 & 4 & 2 \end{array} \right], \\
x_v = \left[ \begin{array}{cccccc} 1 & 1 &  5 & 4 & 4 & 2 \end{array} \right] \]
\endgroup
Now, we have:
\begin{equation}
\textnormal{LiRa}(x_u, x_v) = \log_{10} \frac{ (\frac{1}{2})^6 }{ (\frac{1}{5})^6 } = 2.39
\end{equation}
With twice as much data, the LiRa similarity is twice as high.
Note that, in particular, the maximum LiRa score for any two 
vectors is always attained when the two vectors are equal,
but that the similarity grows as $O(n\log_{10} d)$, where
$n$ is the dimensionality of the input vectors and $d$ is
again the number of discrete rating values. Contrast
this with the Pearson or Cosine similarities,
which will attain a maximum of 1, regardless of the amount of
data observed. 

Like modularity in community detection and the LOD score
in genetic mapping, the LiRA similarity makes assumptions
on the underlying data structure in order to better evaluate
similarity among entities in the data. 

\section{Empirical Evaluation}
\label{sec:empiricalevaluation}
We evaluate the effectiveness of the LiRa similarity in
comparison to other similarity scores in RS data
in two ways: (1) We compare the prediction accuracy of a simple 
kNN method using various similarity scores on real data sets,
and (2) We evaluate the ability of several similarity
scores to distinguish points within the same cluster from
points in different clusters in synthetic data.
Experiments on real data sets show that the LiRa similarity
can detect similar users in a realistic setting with better
accuracy than other scores.
The synthetic data allows us to observe the effect of 
missing entries and dimensionality on similarity computations, 
and to verify that the LiRa score detects users from the same cluster
when a known clustering exists within the data.
\subsection{Data}
As Herlocker et al. note in their overview of methods for evaluating
recommender systems~\cite{herlocker2004evaluating}, there are few publicly available data sets 
that can be used to test hypotheses about RS data, forcing
most research in this field to experiment on the few available data
sources.
Our source of real data are the publicly available MovieLens
data sets, which are among the most often referenced
data sets in RS literature~\cite{bobadilla2013recommender}.
Here, we report results on the 100K and 1M MovieLens
data sets\footnote{\url{http://grouplens.org/datasets/movielens/}}.
For the 100K dataset, we used the u1-u5 \textit{.base}
and \textit{.test} sets when evaluating prediction accuracy. For the
1M dataset, we randomly split the original rating
data into five sets of 80\%/20\% training/test pairs.

In addition to our empirical evaluation on real data, we 
generated a small set of synthetic data sets.
The publicly available data sets we are aware
of are not rich enough to 
examine the effects we are intersted in evaluating -- 
most already contain very high missing rates, 
thus simply deleting existing entries 
to simulate more missing data 
would result in a very limited range of test
cases for experimentation. 
Our experiments on synthetic data give us 
an in-depth view of the effects
of missing entries in RS data. 
\subsection{Experiments on Real Data}
\label{sec:evaluationrealdata}
We first give the pertinent details of our implementation 
of the kNN algorithm, which is used to evaluate
the effectivenes of a similarity score in detecting
users with similar preferences.
For each rating $r_{ui}$ that user $u$ gave to item $i$
in the test set,
we first find at most $k$ nearest neighbors of user $u$
in the training set,
among those who rated item $i$. 
Each neighbor $v$ of $u$ has the property that the similarity
$S(u,v)$ is greater than or equal to $S(u,z)$ for any
other user $z$ in the training set, where $S(u,v)$
 is the similarity score
between $u$ and $v$ in the training set.
The number of neighbors is less than $k$ if less
than $k$ users rated item $i$ in the training set.
Next, the prediction $p_{ui}$ of the rating that
user $u$ gives item $i$ 
is computed by taking an unweighted average of the ratings 
that the neighbors of $u$ have item $i$.

The Root Mean Squared Error (RMSE) ``is perhaps the most popular metric used
in evaluating accuracy of predicted ratings'' ~\cite{gunawardana2015evaluating}.
Another popular measure of prediction accuracy is the Mean Absolute Error (MAE).
The RMSE and 
MAE are defined as follows:
\begingroup\makeatletter\def\f@size{8}\check@mathfonts
\[ RMSE = \sqrt{ \sum_{r_{ui} \in \mathcal{T}} (r_{ui} - p_{ui})^2 }\textnormal{ , }  MAE = \sum_{ r_{ui} \in \mathcal{T}} |r_{ui}- p_{ui}|\]
\endgroup
where $\mathcal{T}$ is the test set of ground truth ratings.
We report both the RMSE and MAE for ratings predicted by the
kNN method for various values of $k$ and several similarity 
scores. 

In addition to the Pearson, Cosine and LiRa scores,
we evaluate the kNN prediction accuracy 
using Patra et al.'s recently proposed BCF score~\cite{patra2015new}.
The Bhattacharyya coefficient for collaborative filtering (BCF) 
attempts to use global item similarities
as weights in local user rating similarity computations,
and was reported to perform well on extremely sparse data sets.  
It is defined as follows:
\begingroup\makeatletter\def\f@size{8}\check@mathfonts
\begin{equation}
\label{eq:bcf}
\textstyle
\textnormal{BCF}(x_i, x_j) = \textnormal{Jacc}(x_i,x_j) + 
\sum_{i \in I_u} 
\sum_{j \in I_v} 
BC(i,j) \textnormal{loc}(x_{ui}, x_{vj})
\end{equation}
\endgroup
where 
\[ BC(i,j) = \sum_{\rho=1}^d \sqrt{\frac{\#\rho_i}{\#i} \frac{\#\rho_j}{\# j} } \]
where $d$ is the number of rating values,
$\#i$ is the number of users that rated item $i$, 
and $\#\rho_i$ is the number of users that rated item $i$ with value $\rho$.  
$I_u$, $I_v$, and $I_{uv}$ are defined as in Section \ref{sec:background}.
Thus $BC$ gives more weight to the local
similarity $\textnormal{loc}(x_{ui},x_{vj})$ 
if items $i$ and $j$ have similar rating distributions
across users in the entire training set.
$\textnormal{loc}(x_{ui},x_{vj})$ is a local similarity measure of the ratings 
that user $u$ gave to item $i$ and user $v$ gave to item $j$.
Of the two \textit{loc} similarity scores defined by Patra et al., 
we chose to use $\textnormal{loc}_{\textnormal{corr}}$, defined as:
\[ \textnormal{loc}_{\textnormal{corr}}(x_u,x_v) = \frac{\left(x_{ui} - \bar{x}_u\right)\left( x_{vj} - \bar{x}_v\right)}{\sigma_u \sigma_v} \]
where $\sigma_u$ is the standard deviation of ratings made by user $u$ and $\bar{x}_u$
is the mean of ratings made by user $u$. In the experimental evaluation of 
Patra et al., $\textnormal{loc}_{\textnormal{corr}}$ 
achieved lower rating prediction error than their other \textit{loc} similarity score.
Jacc$(x_u, x_v)$ is the Jaccard similarity:
\[ \textnormal{Jacc}(x_u, x_v) = \frac{|I_{uv}|}{|I_u|+|I_{v}|} \]
\subsection{kNN Results}
\label{sec:realdataresults}
\begin{figure*}[ht]
  \begin{minipage}[b]{0.5\linewidth}
  \includegraphics[width=\linewidth]{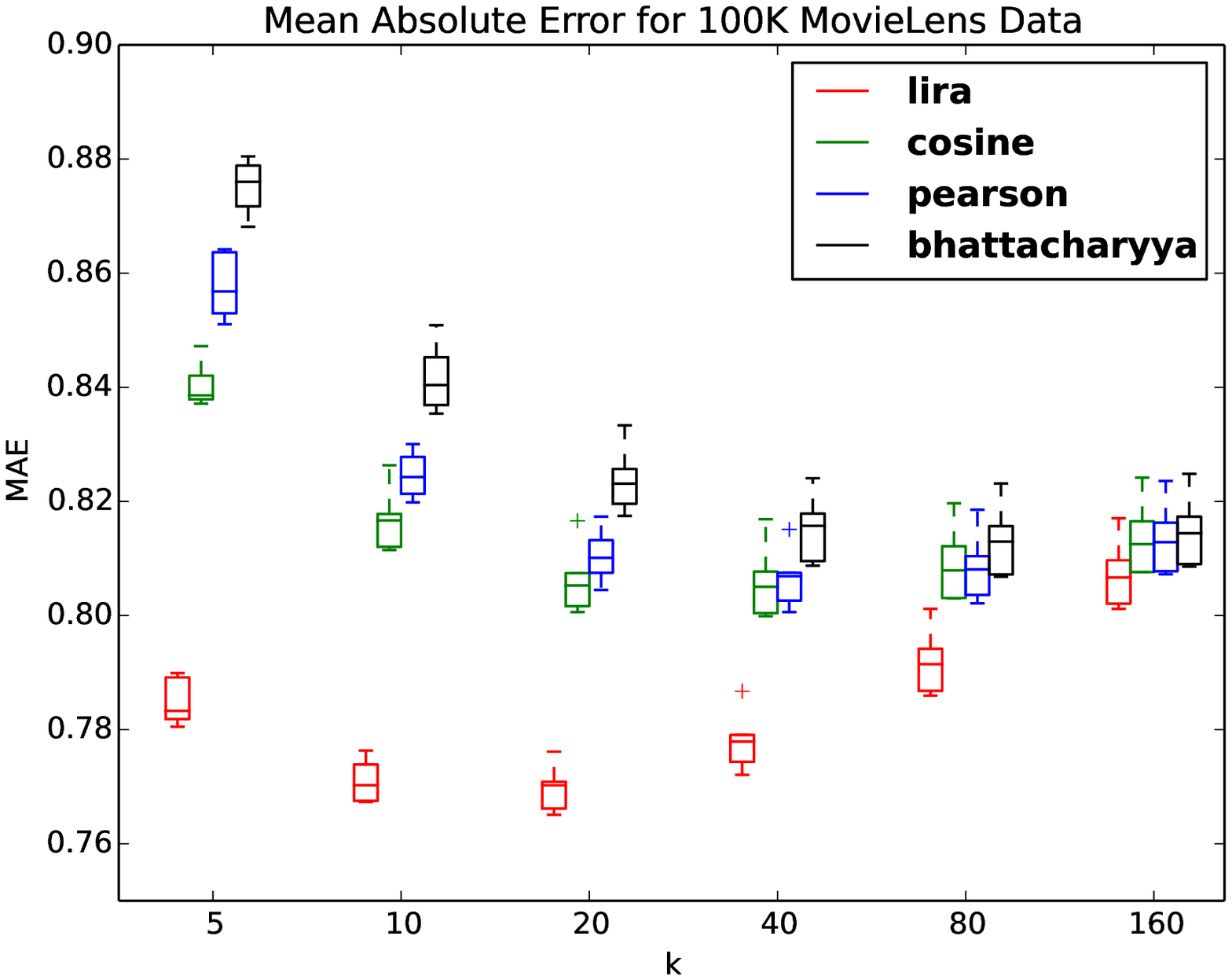}
  \end{minipage}
  \begin{minipage}[b]{0.5\linewidth}
  \includegraphics[width=\linewidth]{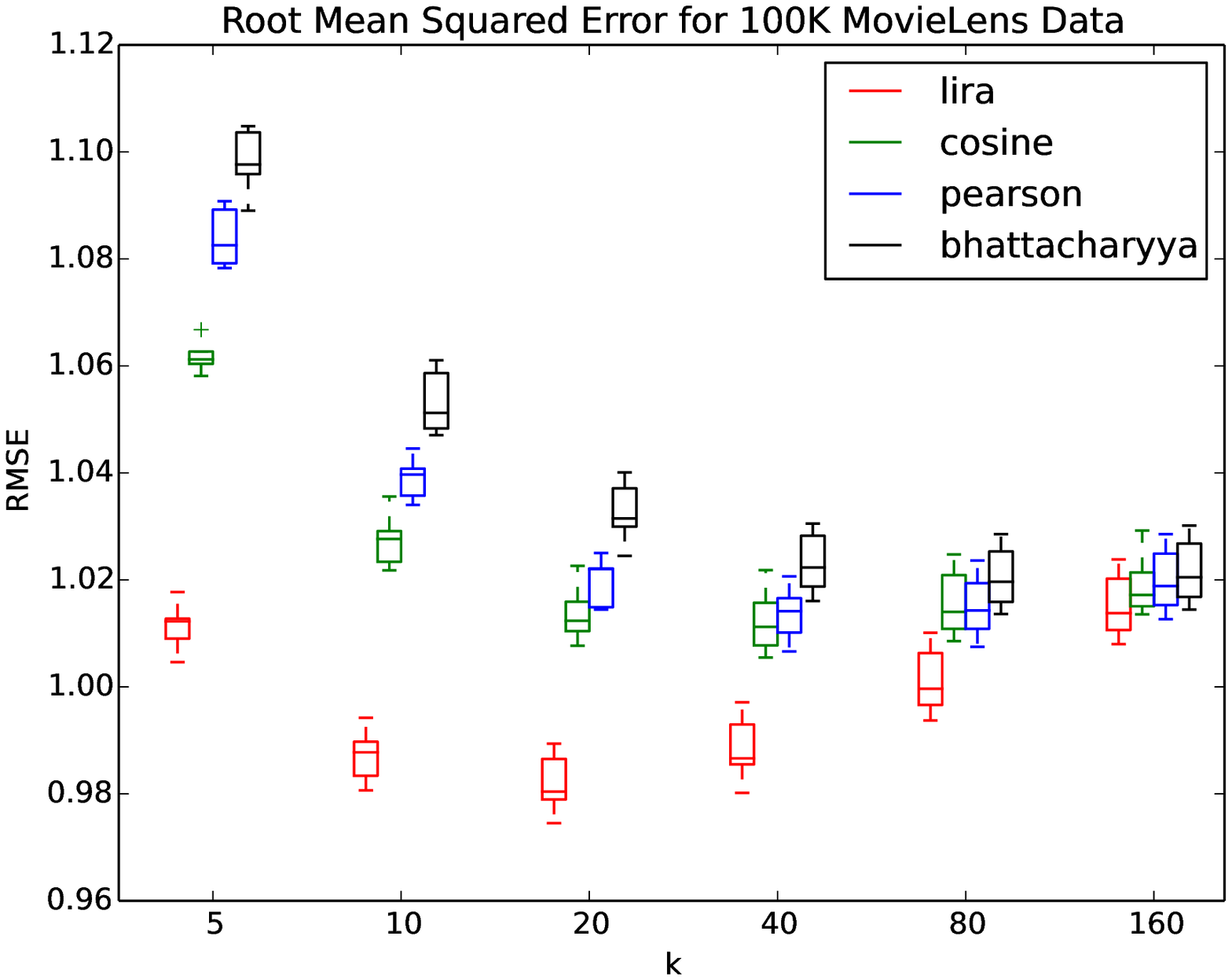}
  \end{minipage}
  \caption{Comparison of prediction accuracy using various similarity scores in kNN show LiRa's better ability to choose relevant neighbors.}
  \label{fig:100kerrors}
\end{figure*}

The MAE and RMSE for kNN prediction on the 
MovieLens 100K data sets
are shown in Figure \ref{fig:100kerrors}, with the MovieLens 1M MAE
results in Figure \ref{fig:mae_1Mdataset}.
Note that the LiRa similarity outperforms other similarity
scores in 
prediction accuracy and for
a wide range of choices for the number of neighbors $k$.
We include both the MAE and RMSE results for the 100K data sets, but 
omit the RMSE for the 1M data sets due to space constraints.
However, the RMSE results for the 1M data sets show the same trend
as is seen in the MAE results -- that is,
LiRa dominates the other similarity scores in accuracy, and the
gap between LiRa and other similarity scores' prediction error widens 
both when we increase the size of the data set, and when we decrease
$k$. 

The kNN curves have the expected shape -- 
at low values of $k$, the data is being under-utilized,
because there are on average 
more than $k$ users who rated item $i$ and are truly similar 
to $u$ in the data, but they are being left out of the
computation of the prediction $p_{ui}$.
At the other extreme, at very high values of $k$,
there are on average less than $k$ truly similar users to $u$
in the data who rated item $i$, and the 
additional users in the neighbor
set are not useful in predicting $u$'s rating of item $i$. 

However, the best 
value of $k$ for LiRa tends to be lower than the best
value of $k$ using other similarity scores, and
the LiRa score outperforms other similarity scores
for all values of $k$ where the neighbor set is not so large that 
it is virtually the same for each similarity score (at $k=160$,
the number of neighbors is 17\% of the 100K training set size,
meaning that for most users, the prediction $p_{ui}$ 
is based on \textit{all} users in the training set who
rated item $i$).
In addition, as $k$ decreases, the gap between LiRa and
the other scores' error widens.
From these observations, we conclude that LiRa
is better able to distinguish between truly similar
and truly dissimilar users; for a given $k$, it 
finds a better set of $k$ neighbors than the other scores,
and as $k$ decreases, it keeps more of the
neighbors that are the better predictors of the rating in
the neighbor set than the other scores. We postulate that
these results demonstrate LiRa's ability to take into account
the amount of data that is being used to evaluate
a similarity score for two users in the data set
in order to make a better determination of similarity.
To test this hypothesis, we next present experiments
on synthetic data, where we can control the missing 
data rate and the ``true" similarity among users. 
\subsection{Experiments on Synthetic Data}
Our goal in generating synthetic data was to evaluate
LiRa's behavior for increasing missing data rates 
in a setting that resembles RS data, but where we can control and understand the 
underlying similarities of users in the data set.
Therefore we use a very simple generative model
to produce 
two clusters $\mathcal{C}_1, \mathcal{C}_2$ of $m/2$ users 
each, where each cluster contains users with similar rating 
patterns on $n$ items and $d$ discrete rating values.
For all experiments in Section \ref{sec:simdataexp}, we
set $m$ to $40$, $d$ to 5, and varied $n$ to 
examine the effects of increasing dimensionality
and user-to-item ratio.
We used the following procedure to generate the two
clusters:
\begin{itemize}
\item[1. ] For each of the two clusters $\mathcal{C}_1, \mathcal{C}_2$, 
and for each of the $n$ items $i$,
randomly choose a $d$-dimensional parameter $\mu_{k i} \in \mathcal{R}^d$,
which defines the multinomial 
distribution $f_{k i }(x = \rho | \mu_{k i} )$ over the $d$ rating
values as: 
\[f_{k i }(x = \rho |\mu_{k i}) = \mu_{k i}(\rho) \]
\item[2. ] For each user $u$ and item $i$ in each cluster $\mathcal{C}_\kappa$, generate rating $x_{ui}$ from 
$f_{k i }(x_{ui} = \rho| \mu_{k i})$
\end{itemize}
Thus users from the same cluster will tend to 
have similar rating patterns.
To illustrate our simple model we provide the following
example: suppose we set $m=40$, $n=2$, and $d=5$.
Our simulation produces the following $5$-dimensional parameters 
$\mu_{k i} \in \mathcal{R}^5$:
\begingroup\makeatletter\def\f@size{8}\check@mathfonts
\[ \mu_{1 1} = \left[ 0.55, 0.09, 0.25, 0.01, 0.10 \right], 
\mu_{1 2} = \left[ 0.34, 0.33, 0.29, 0.03, 0.01 \right] \]
\[ \mu_{2 1} = \left[ 0.17, 0.08, 0.12, 0.33, 0.30 \right], \mu_{2 2} = \left[ 0.04, 0.25, 0.47, 0.07, 0.18 \right] \]
\endgroup
Therefore users in cluster $\mathcal{C}_1$ tend to rate item 1 
about half the time with a value of 1 (since $\mu_{1 1}(1) = 0.55$) 
about a quarter of the time with a value of 3 ($\mu_{1 1}(3) = 0.25$), and much less
often with a value of 2, 4, or 5.
The same users in cluster 1 tend to rate item 2 with a value of either 1,2, or 3,
and much less often with a value of 4 or 5.
Therefore, one likely user $u$ from $C_1$ can be represented
by the vector $x_{u} = \left[ 1, 2\right] $, whereas
a likely user from $C_2$ is $x_v = \left[ 4, 2\right]$.

Note that the parameters $\mu_{k i}$
are generated at random, but sum to one and are consistent within
a cluster, ensuring that users from the same cluster
rate items with the same patterns. Therefore, we expect that 
the similarity between two users within the same cluster should
be high when compared to the similarity of two users
in different clusters. However, we did not explicitly
generate the idealized clusters that make up the model
used in our LiRa score, to show that the oversimplified
model used by LiRa is nevertheless enough to capture much of
the intra-cluster similarity and inter-cluster dissimilarity.

We quantify a similarity score's ability to resolve 
two users in the same cluster from two users
in different clusters
with a quantity we call the score's \textit{resolution}.
The resolution of a score $S$ is defined as the mean
of $S(x_u,x_v)$ for all points $x_u$ and $x_v$ within the same 
cluster minus the mean of $S(x_u, x_w)$ for
all points $x_u$ and $x_w$ in different clusters.
In experiments we set $S$ to the LiRa,
Pearson, Cosine, and Bhattacharyya similarity scores, for 
increasing values of the dimensionality $n$.
A positive resolution value indicates greater average intra- than inter-
cluster similarity, meaning that a score $S$ is greater
for points in the same cluster than points
in different clusters on average.
To additionally observe the effect of missing data on the similarity
scores, we randomly deleted an increasing fraction of the ratings $x_{ui}$.
Thus the expected number of co-observed entries decreases
with the missing rate.
\subsection{Results on Synthetic Data}
\label{sec:simdataexp}
\begin{figure*}[ht] 
  \begin{minipage}[b]{0.45\linewidth}
  \centering
  \includegraphics[width=\linewidth]{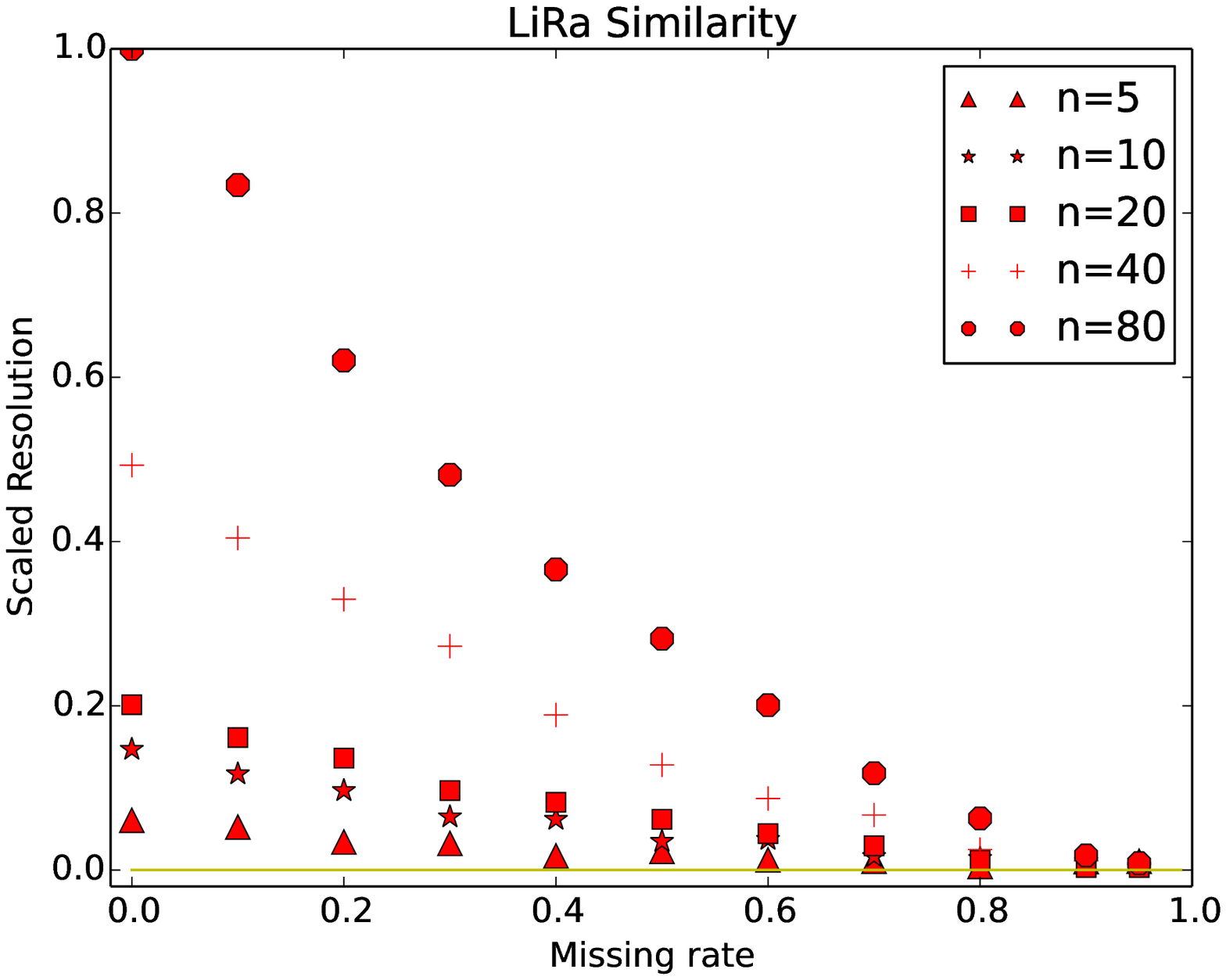}
  \end{minipage}
  \begin{minipage}[b]{0.45\linewidth}
  \centering  
  \includegraphics[width=\linewidth]{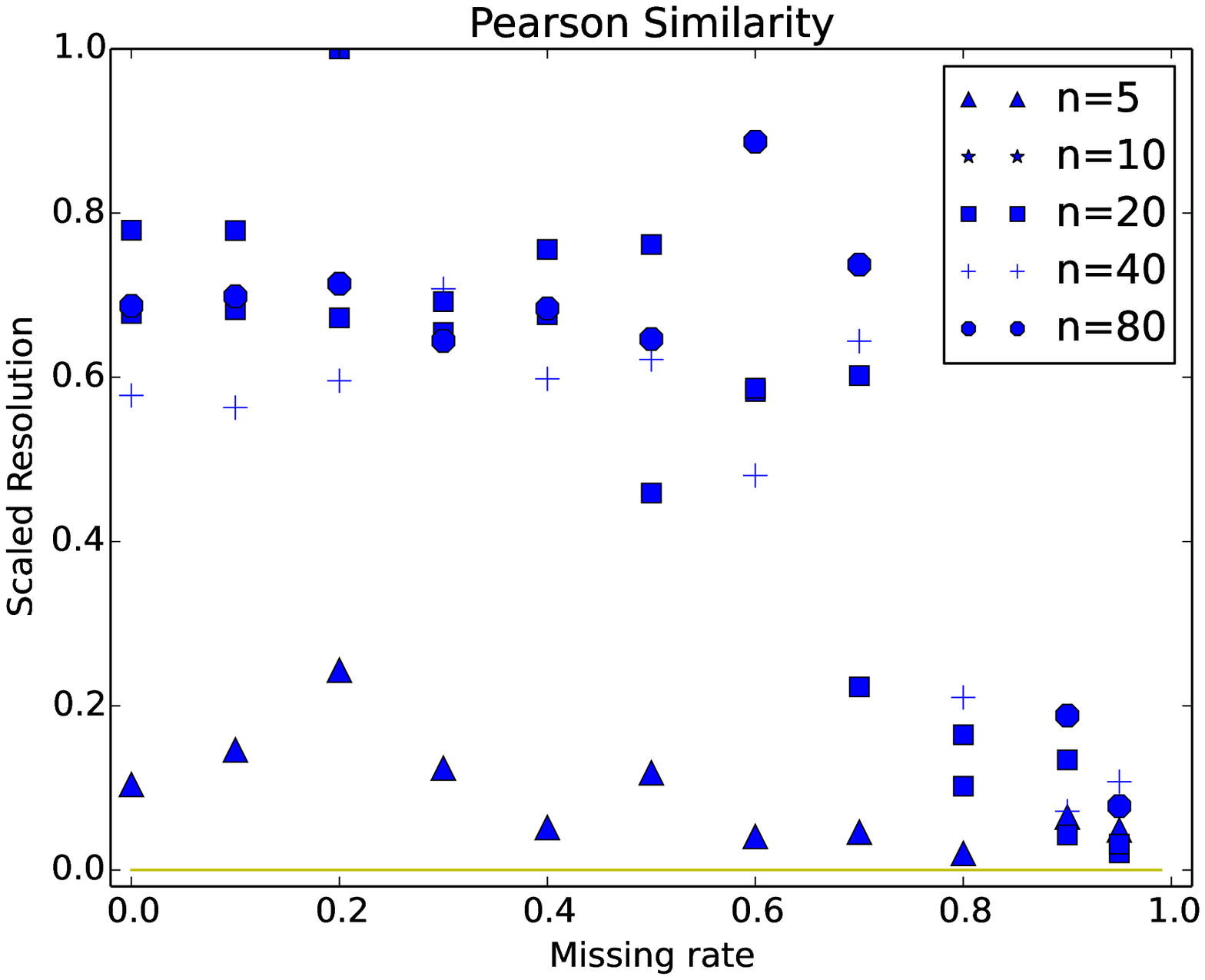}
  \end{minipage}
  \newline
  \begin{minipage}[b]{0.45\linewidth}
  \centering
  \includegraphics[width=\linewidth]{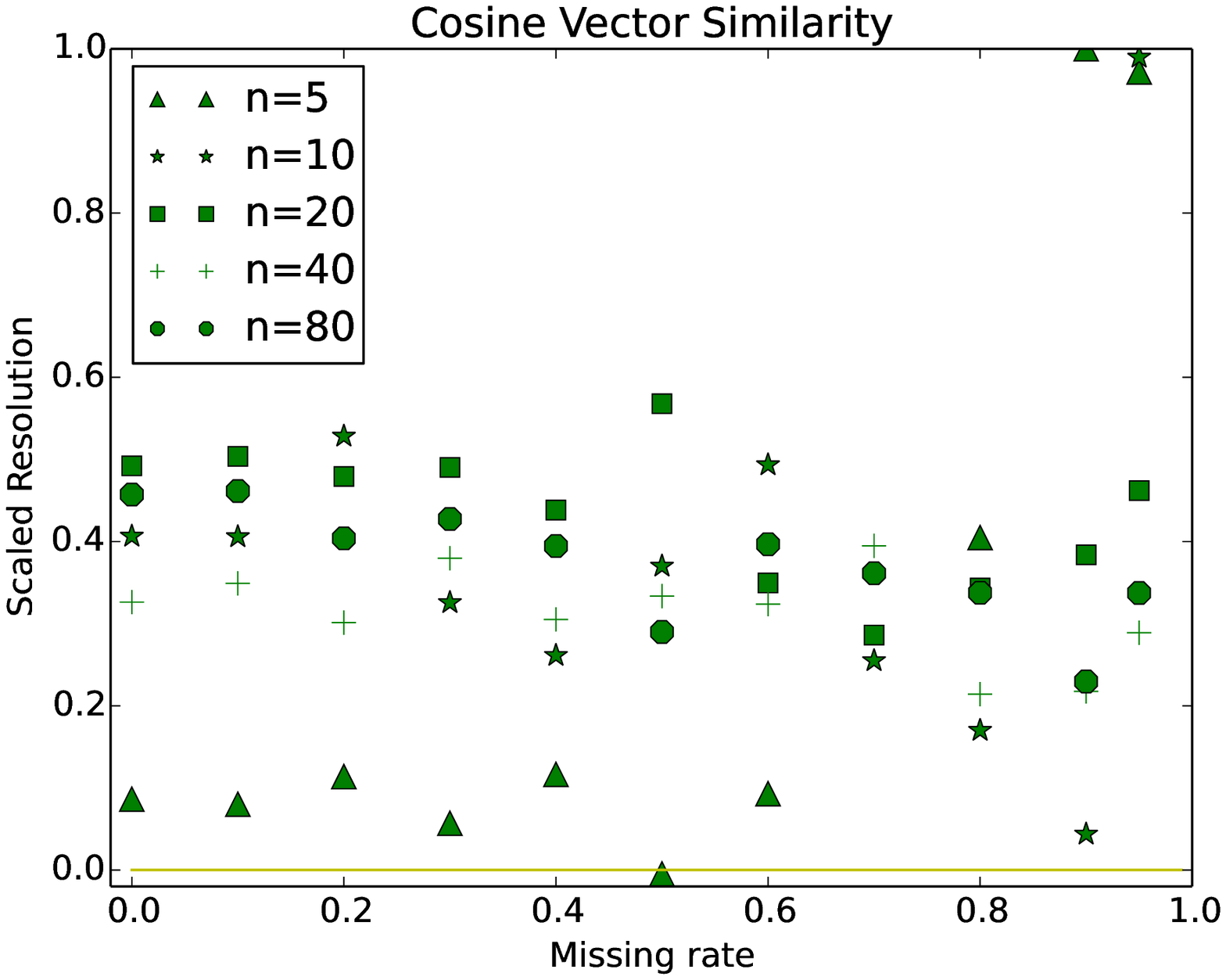}
  \end{minipage}
  \begin{minipage}[b]{0.45\linewidth}
  \centering
  \includegraphics[width=\linewidth]{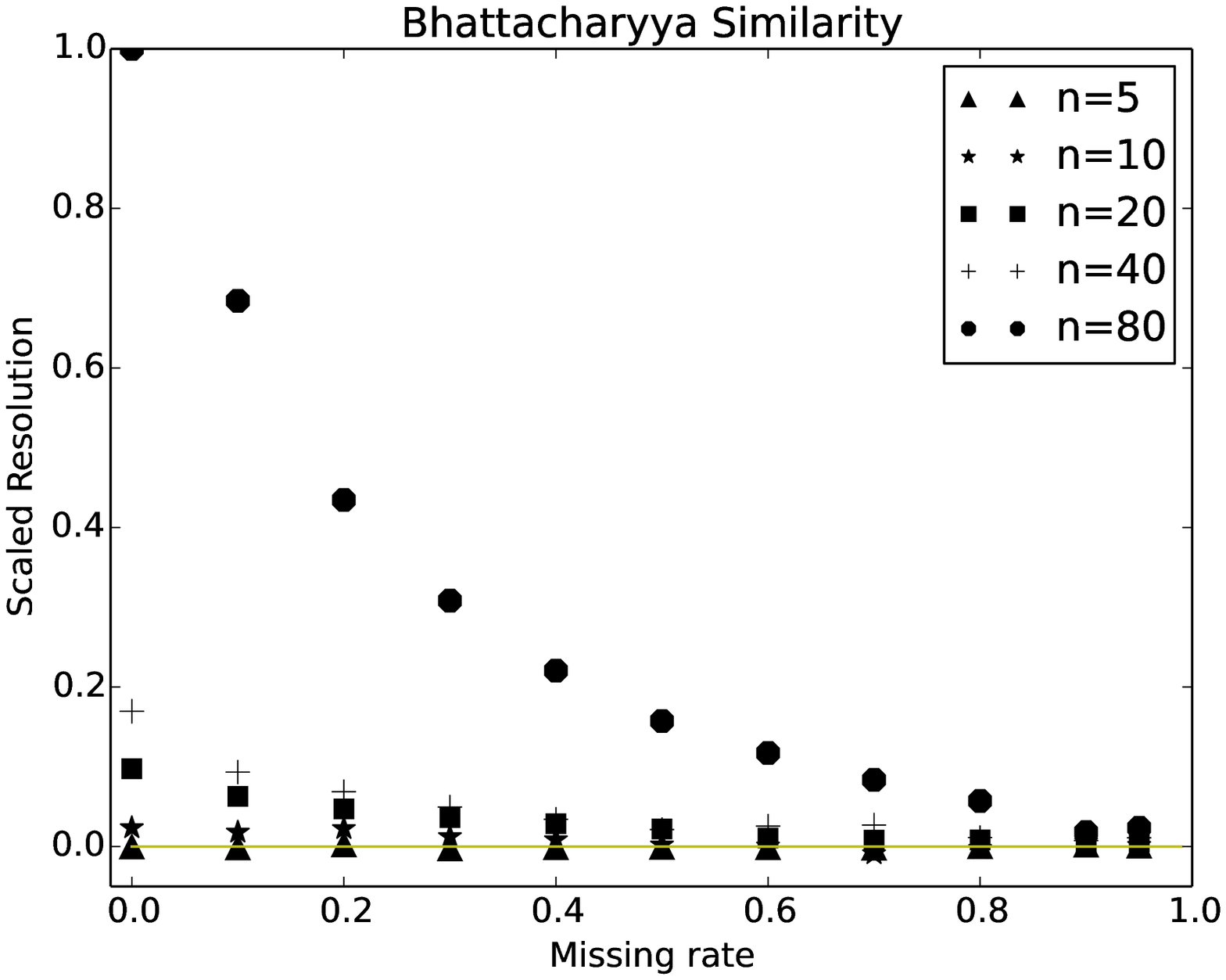}
  \end{minipage}
  \caption{Similarity resolution (higher is better) as a function of missing data rate plotted 
  for four similarity scores: LiRa (upper left), Pearson (upper right),
  Cosine (lower left), and Bhattacharyya (lower right). Resolution indicates a score's ability
  to differentiate a pair of points in the same cluster from a pair of points in different clusters.
  Cosine and Pearson scores do not improve in resolution with more data availability.}
  \label{fig:simdiff_all}
\end{figure*}

The scaled resolution is plotted in 
Figure \ref{fig:simdiff_all} 
for the LiRa (red), Pearson (blue), Cosine (green)
 and Bhattacharrya (black) similarities.
Each marker in each plot corresponds to a different dimensionality
$n$, where $n$ increases from 5 to 80, doubling each time,
and the missing rate increases from 0.1 to 0.9 in increments of 0.1,
with an additional point at 0.95.
We scaled each score's resolution by dividing by the maximum-magnitude resolution achieved
by that score in the experiments.
Therefore, a scaled resolution of 1 indicates the missing rate and dimensionality
with the
highest-magnitude resolution over all missing rates and dimensionalities,
and a magnitude less than one tells us what
fraction of the maximum-magnitude resolution
was achieved at a particular dimensionality and missing rate. 

We observe 
that the resolution of the LiRa score
is greater as we decrease missing data
and increase the dimensionality.
In addition, the LiRa resolution is positive
for all values of the missing rate and 
all dimensionalities (the minimum of LiRa
resolutions was 0.051), indicating
greater average intra- than inter-cluster LiRa similarity.
The greater magnitude of the resolution 
in the presence of more data shows the 
ability of LiRa to make
stronger claims about similarity 
as the number of co-observed entries
in two discrete-valued vectors increases. 
More data comes in the form of a lower
missing data rate, but also increased dimensionality,
because there will be more expected co-observed entries
between two vectors when the dimensionality
is higher. 
 
Contrast this with the Pearson and Cosine similarity
scores, where dimensionality and missing data
have virtually no effect on the resolution, except 
that high values of missing rates tend to lower
the Pearson resolution dramatically. 
The resolution is positive for most values of
the dimensionality and missing rates for Cosine, 
and all values of dimensionality and missing rates
for Pearson, but increasing
the amount available data does not improve the ability
of Cosine or Pearson to resolve similar from dissimilar
users. 
The fact that low dimensionalities and higher missing
rates often yield a higher resolution than higher
 dimensionalities and lower missing rates shows theses scores' inability
to make use of more data for more accurate similarity judgments.
For the most part, the Bhattacharyya resolution tends to increase
with increasing dimensionality and decreasing missing rates,
also indicating a greater difference between intra- and inter-
cluster scores,
but there are instances when this is not the case. 

\begin{figure}
  \includegraphics[scale=0.45]{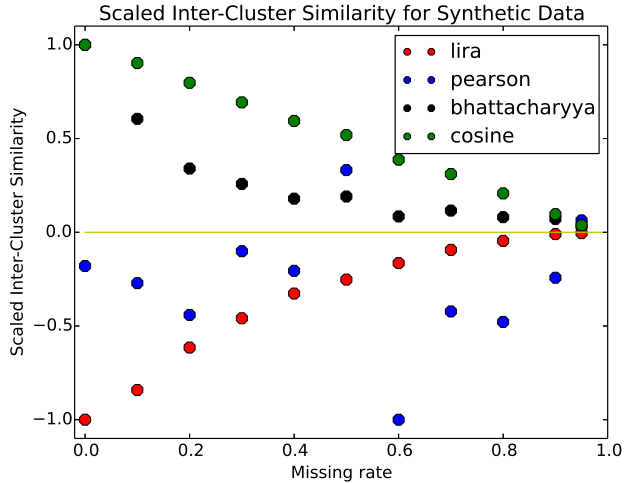}
  \caption{The inter-cluster similarity of various similarity
  scores on synthetic data. LiRa is the only score which   
  decreases with decreasing missing rate, indicating greater
  ability to differentiate points from different clusters when more data
  is available. }
  \label{fig:c12sim}
\end{figure}

We conclude this section with a discussion of
Figure \ref{fig:c12sim}, which plots the scaled average
inter-cluster similarity score across missing rates for
the four tested scores, with a fixed 
dimension of $n=80$.
Scaling was again done by dividing each average 
inter-cluster similarity by the magnitude of the 
greatest-magnitude average inter-cluster similarity that occurred
over all missing rates.
This way, we can see 
how inter-cluster similarity changes with the missing rate,
for all scores on the same scale.

Recall that a negative Pearson or LiRa value
indicates dissimilarity in some way. 
A negative Pearson score indicates anti-correlated co-observed
entries.
A negative LiRa score indicates a greater chance
that the data in co-observed entries 
was generated by chance, rather than that the data
comes from two vectors in the same cluster.
In Bhattacharyya, a negative score also indicates
anti-correlation in the co-observed entries,
but weighted by item similarity and shifted
by the Jaccard similarity, making it harder to interpret.
Cosine is restricted to positive values in this
setting, because
all vector entries were positive.

We observe that LiRa and Pearson appear to be 
better able to indicate dissimilarity,
as their scaled average inter-cluster values
are negative for most and all 
missing rates, respectively. However, LiRa again
makes use of more data to make a stronger claim
about dissimilarity, giving greater-magnitude negative
values when more data is observed.
Both Cosine and Bhattacharyya remain positive,
indicating some similarity between points from
different clusters, and actually increase in magnitude
as more data becomes available, scoring
two points from different clusters higher at lower
missing rates. 
Based on the MovieLens results, this may also mean 
 that the real data contains user clusters, and the 
 Bhattacharyya score is too high for users from 
 different clusters. Thus it may choose
 sub-optimal neighbors in the kNN method in this case.

Based on our results from these synthetic data experiments, 
we make two concluding remarks about the results on
real data in Section \ref{sec:realdataresults}.
First, the high performance of LiRa on the MovieLens data can 
be explained by its dependence on not only the rating patterns in co-observed
entries of user rating vectors, but also on
the amount of data that is available to make the similarity computation.
Second, based on the Bhattacharyya results, we believe
a promising future research direction is to further
investigate what type of underlying cluster structure
exists in RS data. The intuition behind the Bhattacharyya similarity is
 that a higher similarity between two items in the set $I_{uv}$,
 defined by the difference between their rating distributions
 across the entire data set, should contribute
 a higher weight to the difference in these item ratings.
 Each of the $|I_{uv}|^2$ pairs of co-rated items
 $(i,j)$ contributes to the similarity score.
By contrast, LiRa does not consider item similarity in the
 computation of user similarity, and only examines the 
 differences in user ratings of the $|I_{uv}|$ co-rated items. 
A clustering structure such as that assumed 
by LiRa may indeed exist in real-world
data, and perhaps the Bhattacharyya score is 
not well suited to this setting, where it can be high
for users from different clusters, despite its
ability to give greater scores to users from
the same cluster with more data.

\section{Related Work}
\label{sec:relatedwork}
Early studies ~\cite{sarwar2001item, herlocker2002knnempirical, desrosiers2011comprehensive} 
consistently rate the Pearson, Cosine Vector,
or slight variants as the superior similarity scores for RS data.
Several recent studies~\cite{ahn2008new, patra2015new} focus on the cold start problem, 
in which extremely few ratings
are available for a new user, making it difficult to determine
her similarity to other users with traditional similarity
scores.

The PIP heurisitic was introduced 
by Ahn~\cite{ahn2008new} 
to address the cold-start problem, 
but was shown to perform comparably to traditional
similarity scores such as Pearson's correlation coefficient in
non-cold start settings, and was outperformed in terms of
rating prediction
accuracy by the Bhattacharyya coefficient for collaborative
filtering in cold start settings with extremely few ratings.
Patra et al.'s 
Bhattacharyya coefficient for collaborative filtering,
defined in Section \ref{sec:empiricalevaluation}, 
takes into account item similarity as a weighting scheme for user similarity.
It was developed for the extremely
sparse setting, 
thus Patra et al.'s empirical evaluation of its
use in rating prediction 
accuracy is restricted to data sets where the
missing rate is much higher than even the already sparse
MovieLens datasets, and is better suited
for the cold-start setting.
The Bhattacharyya coefficient also
is more computationally expensive than our LiRa score, 
as it requires all-to-all user-to-user 
as well as all-to-all item-to-item
similarity computations.

In the context of statistical text analysis, 
Dunning~\cite{dunning1993accurate} makes a strong case
for the use of a log-likelihood ratio to examine the
statistical significance of word or bigram frequencies
in sparse data.
The hypotheses in the LiRa ratio express 
whether users are or 
are not similar, based on the assumption
that user similarity is reflected by the probability
of two users' ratings being distributed according
to the clustering model presented in Section \ref{sec:liradef}.
These are different from the hypotheses of the dimensionality
of the underlying parameter space as defined by Dunning. 
Dunning's work has been adapted to RS data 
in several ways
and has been shown to enhance the performance of 
recommender systems in an industrial
setting~\cite{casinellievaluating, paraschakiscomparative}.
However, we emphasize that in the applied versions of Dunning's
likelihood ratio to RS data,
the ratio is used either as a filter for finding relevant 
items to use in similarity computations, or a weighting term
in the rating prediction phase,
and has not been developed into a similarity score. 
Our method also slightly resembles
Jojic et al.'s~\cite{jojic2011probabilistic}
item similarity score, where
similarity is determined 
by comparing
the number of users that like two items to the number
expected by chance.
However, their method is limited to binary like/dislike 
data,
treats dislikes the same as missing entries 
in the
item similarity computation, and was used
in combination with additional heuristics
to determine whether a user will like a particular item.

The intuition that inherent clusters of users exist
in RS data has been explored by
several clustering methods which were
developed to improve prediction accuracy in RS data.
Sarwar et al.~\cite{sarwar2002recommender} used clustering to improve scalability
by first partitioning the users into clusters, then making a
prediction based on averaging ratings from members of the same cluster,
allowing for less computation time than a kNN method
on the MoveLens 100K data set.
Similarly, Xue et al. introduce a $k$-means clustering phase prior 
to prediction~\cite{xue2005scalable}, and predict a rating for a user $u$
by choosing $k$ neighbors out of the clusters with 
\textit{representatives} that score highly with $u$.
Rashid et al.~\cite{al2006clustknn} incorporate bisecting $k$-means clustering ``to increase efficiency and
scalability while maintaining good recommendation quality" in
their ClustKNN algorithm.
Das et al.~\cite{das2014clustering, das2015iterative}
use a DBSCAN-based algorithm to improve kNN prediction accuracy.
An extensive experimental study of the effectiveness of 
various centroid selection 
methods for the $k$-means algorithm when used as a pre-processing 
step in recommendation systems is presented by Zahra et al.~\cite{zahra2015novel},
who conclude that although many approaches improve prediction accuracy and
efficiency,
no algorithm is ``a panacea" across all data sets.
Nonetheless, the results of clustering-based approaches in rating prediction are encouraging,
as they show promise in improving both accuracy and scalability of
recommender systems.

\section{Conclusion}
We have introduced the LiRa similarity score for discrete-valued,
sparse, and high-dimensional data, typical of the
RS domain. We have shown through empirical evaluations on both 
real and synthetic data that LiRa's assumptions about a clustering
model of users makes it a good indicator of user similarity,
and that it outperforms other measures in this capacity.
An exciting area to focus our future research
is exploring how to devise a better model of clustering structure
within RS data in order to improve prediction accuracy of collaborative
filtering methods.
Another possible research direction is to develop fast clustering methods 
that use LiRa as a similarity score to improve the scalability of
user-based collaborative filtering.
\section*{Acknowledgment}
This work is supported by the Applied Mathematics Program of the DOE Office of Advanced Scientific Computing Research under contract number DE-AC02-05CH11231 and by NSF Award CCF-1637564.

%

\bibliographystyle{abbrv}
\bibliography{references}  
%
%

\end{document}